\documentclass[reprint,superscriptaddress,amsmath,amssymb,aps,prl]{revtex4-1}

\usepackage{graphicx}
\usepackage{bm}
\usepackage{color}
\usepackage{graphicx}
\usepackage{subfigure}
\usepackage{hyperref}

\newcommand{\ud}{\mathrm{d}}
\usepackage{amsmath}

\begin{document}

\title{Modelling Elastically-Mediated Liquid-Liquid Phase Separation}

\author{Xuefeng Wei}
\thanks{X. W. and J. Z. contributed equally to this work. }
\affiliation{CAS Key Laboratory of Theoretical Physics, Institute of Theoretical
Physics,Chinese Academy of Sciences, Beijing 100190, China}
\affiliation{School of Physical Sciences, University of Chinese Academy of Sciences, China}

\author{Jiajia Zhou}
\thanks{X. W. and J. Z. contributed equally to this work. }
\affiliation{Key Laboratory of Bio-Inspired Smart Interfacial Science and Technology of Ministry of Education, School of Chemistry, Beihang University, Beijing 100191, China}
\affiliation{Center of Soft Matter Physics and Its Applications, Beihang University, Beijing 100191, China}

\author{Yanting Wang}
\affiliation{CAS Key Laboratory of Theoretical Physics, Institute of Theoretical
Physics,Chinese Academy of Sciences, Beijing 100190, China}

\author{Fanlong Meng}
\email{Corresponding author: fanlong.meng@itp.ac.cn}
\affiliation{CAS Key Laboratory of Theoretical Physics, Institute of Theoretical
Physics,Chinese Academy of Sciences, Beijing 100190, China}

\begin{abstract}
We propose a continuum theory of the liquid-liquid phase separation in an elastic network where phase-separated microscopic droplets rich in one fluid component can form as an interplay of fluids mixing, droplet nucleation, network deformation, thermodynamic fluctuation, \emph{etc}.
We find that the size of the phase separated droplets decreases with the shear modulus of the elastic network in the form of $\sim[\mathrm{modulus}]^{-1/3}$ and the number density of the droplet increases almost linearly with the shear modulus $\sim[\mathrm{modulus}]$,
which are verified by the experimental observations.
Phase diagrams in the space of (fluid constitution, mixture interaction, network modulus) are provided, which can help to understand similar phase separations in biological cells and also to guide fabrications of synthetic cells with desired phase properties.
\end{abstract}

\maketitle

Membraneless compartments/organelles in cells are often supramolecular assemblies composed of proteins, nucleic acids, and other molecules~\cite{Schmidt2016, Uversky2017, Rai2018, Marnik2019}.
Examples include nucleolus in the nucleus, stress granules, centrosomes in the cytoplasm, \emph{etc.}, and they usually provide physical constraints for specific biochemical reactions.
After identifying the liquid-like features of some membraneless compartments represented by P granules~\cite{Brangwynne2009}, liquid-liquid phase separation was proposed as a plausible mechanism for the formation of such membraneless compartments~\cite{Hyman2012, Lee2013, Hyman2014, Boeynaems2018, Alberti2019}.
Differently from typical liquid-liquid phase separation in one component system~\cite{Franzese2001, Katayama2000, Tanaka2000} or multi-component mixtures~\cite{Huang1995, MaoSheng2019}
where the mixing energy of liquids is deterministic of whether phase separation can occur or not,
the case in cells is usually more complicated due to the elastic constraint by cytoskeleton~\cite{Amos, Howard}.
Thus, exploring the role of the elastic network in liquid-liquid phase separation will be important in understanding and predicting the properties of the phase-separated products.

Recent experimental studies~\cite{Style2018, Rosowski2020} found that the liquid-liquid phase separation of a fluorinated oil--silicone oil mixture in a silicone polymer network can occur with the formation of droplets rich in one fluid component (fluorinated oil); with the increase of the elastic modulus of the network, the size of the droplets decreases while the number density increases.
The critical concentration of the fluorinated oil for its condensation (droplet formation) in the elastic network has been related with the stiffness of the network by thermodynamic arguments~\cite{Kim2020, Rosowski2020a} and the dynamics of droplets due to stiffness gradient is discussed \cite{Vidal-Henriquez2020, Kothari2020}, while how to theoretically characterize the relation between the droplet properties (size, number density, \emph{etc}.) and the network elasticity remains unclear.
The main difficulty lies in how to deal with the system complexity incorporating mixing of different fluids, droplet nucleation in a polymer network, thermodynamic fluctuation, \emph{etc}.

\begin{figure}[ht]
\centering
\includegraphics[width=0.89278\columnwidth]{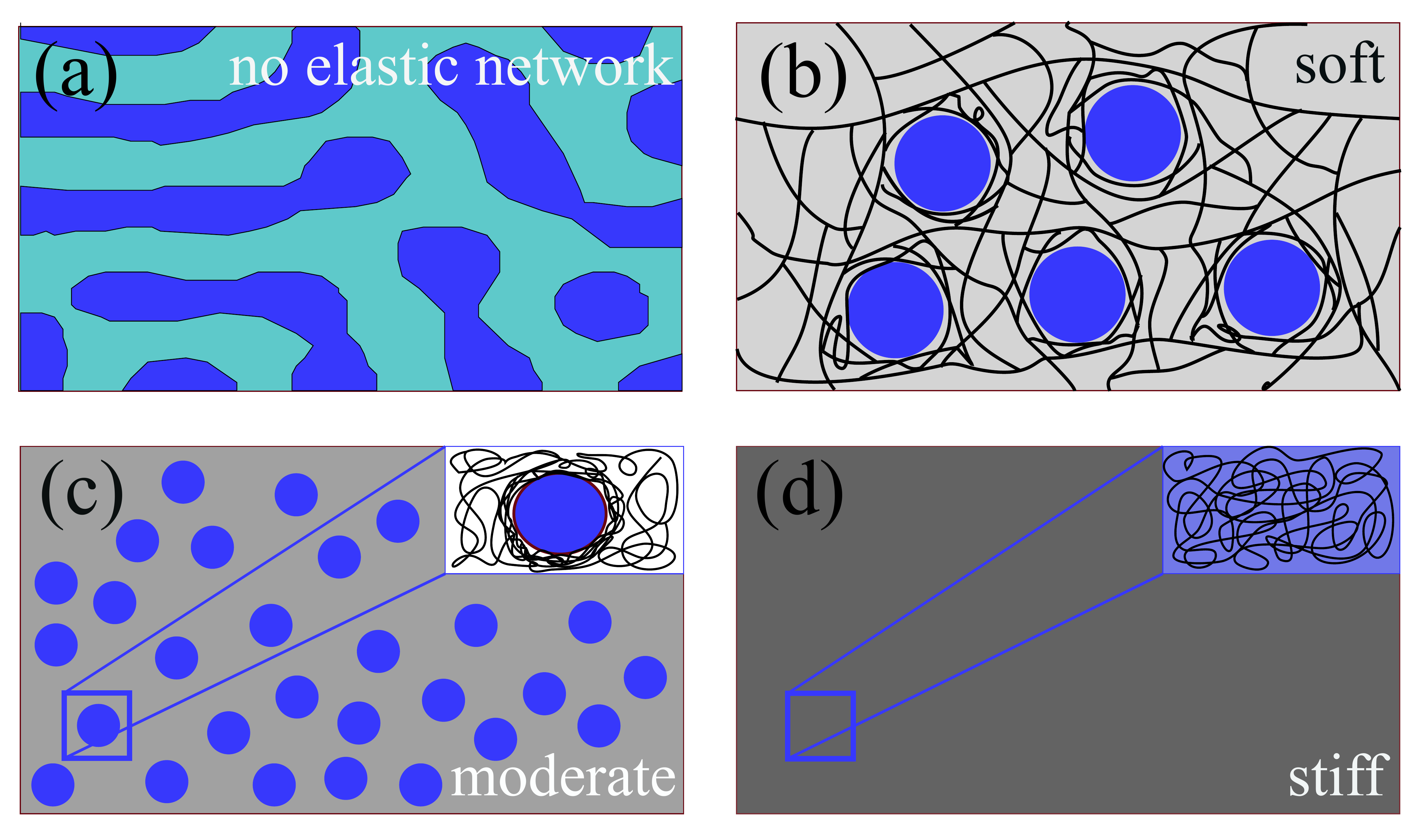}
\caption{(a) Liquid-liquid phase separation in a free space (without elastic network). (b) A few large droplets form in a soft elastic network (small shear modulus). (c) Many small droplets form in a moderate elastic network (moderate shear modulus). (d) Inhibition of phase separation in a stiff elastic network (large shear modulus).}
\label{figl}
\end{figure}

In this work, we treat the liquid-liquid phase separation of an $A$/$B$ fluid mixture in an elastic polymer network.
By constructing a free energy model of the system, which incorporates the mixing entropy, the Flory-Huggins type interaction among fluid mixture, the surface energy of phase-separated droplets, and the elastic energy of the droplets-deforming polymer network,
we discuss thoroughly how network elasticity determines the conditions of the phase separation and also the properties of phase-separated products.

When liquid-liquid phase separation occurs, droplets rich in one fluid component will form, and the number density and the radius of the droplets are denoted as $n$ and $R$, respectively; then the volume fraction of the droplet is equal to $\nu=4\pi n R^3/3$.
Here we assume the droplets are monodisperse in size, following the experimental observations~\cite{Style2018}, and such monodispersity results mainly from the elastic constraint by the polymer network.
Considering the volume conservation of $A$ component, there is the relation between the volume fraction of $A$ component in the droplets $\phi_{A}^{\rm d}$ and in the bulk $\phi_{A}^{\rm b}$ as:
\begin{equation}
  \label{eq:vol_cons}
  \nu \phi_{A}^{\rm d} + (1-\nu) \phi_{A}^{\rm b} = \phi_A^{0} \, ,
\end{equation}
where $\phi_A^{0}$ is the overall volume fraction of $A$ component as in the prepared state (before phase separation).
The free energy density of the system can be expressed as
\begin{eqnarray}
  && \mathcal{F}(\phi_{A}^{\rm d},\phi_{A}^{\rm b}, \nu, R) = \nu f_{\rm mix}(\phi_A^{\rm d}) + (1-\nu) f_{\rm mix}(\phi_A^{\rm b}) \nonumber \\
  && \quad + 3 \nu \frac{\gamma}{R} + \nu f_{\rm el}(R)- \xi \left[ \nu \phi_A^{\rm d} + (1-\nu) \phi_A^{\rm b} - \phi_A^0 \right] .
     \label{f_energy}
\end{eqnarray}
The first and the second term on the right hand side of Eq.~(\ref{f_energy}) denote the mixing free energy density of two fluids (without elastic constraint) inside and outside of the droplets, respectively.
For this free energy density, $f_{\mathrm{mix}}(\phi_{A})$,
we adopt the typical Flory-Huggins form~\cite{DoiSoft},
which consists of the mixing entropy and the interaction between the fluids, as: $f_{\mathrm{mix}}(\phi_A)=\frac{k_{\mathrm{B}}T}{v_A}(\phi_{A} \ln\phi_{A}+\frac{v_A}{v_B}\phi_{B}\ln\phi_{B}+\chi \phi_{A}\phi_{B})$,
where $k_{\rm B}$ is Boltzmann constant, $T$ is temperature, $v_A$ and $v_B$ denote the volume of one \emph{A}- and one \emph{B}- liquid molecule, respectively, $\phi_{B}=1-\phi_{A}$ is the volume fraction of \emph{B}-component and $\chi$ is the Flory-Huggins parameter characterizing the interaction between \emph{A}- and \emph{B}- liquids.
Note that we treat the network component the same as \emph{B}-component in order to compare with relevant experiments in the later discussion, while it is straightforward to adapt the theory to the case where the network component is different from \emph{A}- or \emph{B}-component.
The third term in Eq.~(\ref{f_energy}) denotes the interfacial energy density of the droplets where $\gamma$ is the surface tension.
The fourth term in Eq.~(\ref{f_energy}) denotes the elastic energy induced by the droplets deforming the elastic network~\cite{ZhuJian2011, Hutchens2016}.
The explicit expression of $f_{\rm el}$ is taken as: $f_{\rm el}(R) = 3 \, [ 1 -  \left(R_0/R\right)^3 ] \int_1^{R/R_0} \lambda^2 W(\lambda) / (\lambda^3-1)^2 \ud \lambda$, where $R_0\simeq (k_{\mathrm{B}}T/G)^{1/3}$~\cite{de1979} denotes the mesh size of the network with $G$ as the shear modulus of the elastic network, \emph{e.g.}, $R_0\simeq 0.01 \, \mathrm{\mu m}$
for $G \simeq 5 \, \mathrm{kPa}$.
The function $W(\lambda)$ in this elastic term represents the elastic energy density of an inflated spherical shell (like a spherical balloon), for which we take the Gent model~\cite{gent1996}, \emph{i.e.},
$W(\lambda)=-\frac{1}{2} G J_{\mathrm{m}} \ln [1 - {J(\lambda)}/{J_{m}}]$,
where $J(\lambda)=2\lambda^2+\lambda^{-4}-3$ with $\lambda$ as the stretch ratio in the radial direction and $J_{\mathrm{m}}$ is a phenomenological parameter characterizing the finite stretchability of the material.
Note that if the deformation is small, \emph{i.e.}, $\lambda\rightarrow 1$,  the Gent model reduces to the neo-Hookean model, with $W(\lambda)=\frac{1}{2} GJ(\lambda)$.
The last term in Eq.~(\ref{f_energy}) comes from the constraint in Eq.~(\ref{eq:vol_cons}) with $\xi$ as the Lagrangian multiplier.

\paragraph{Chemical and mechanical balance.}
By optimizing the total energy density with respect to the volume fraction of \emph{A}- component in the phase-separated droplets $\phi_A^{\rm d}$ and in the bulk $\phi_A^{\rm b}$, respectively,
$\partial \mathcal{F}/\partial \phi_A^{\rm d} =
\partial \mathcal{F}/\partial \phi_A^{\rm b} = 0$,
one identifies $\xi$ as the chemical potential,
\begin{equation}
  \label{chemical}
  \xi = f'_{\rm mix}(\phi_A^{\rm d}) = f'_{\rm mix}(\phi_A^{\rm b}).
\end{equation}
In other words, the chemical potential of \emph{A}- (\emph{B}-) component in the phase separated droplets and in the bulk should be equal.
By optimizing the energy density with respect to the volume fraction $\nu$, \emph{i.e.}, $\partial \mathcal{F}/\partial \nu= 0$, there is the relation:  $f_{\rm mix}(\phi_A^{\rm d}) - f_{\rm mix}(\phi_A^{\rm b}) - \xi (\phi_A^{\rm d} - \phi_A^{\rm b}) + 3\gamma/R + f_{\rm el}(R)= 0$.
With the substitution of the chemical potential $\xi = f'_{\rm mix}(\phi_A^{\rm d}) = f'_{\rm mix}(\phi_A^{\rm b})$, the pressure difference between the droplet and the bulk, $\Delta P$, should obey:
\begin{equation}
  \label{mechanic}
  \Delta P = \Pi(\phi_A^{\rm d})-\Pi(\phi_A^{\rm b}) = \frac{3\gamma}{R}+ f_{\rm el}(R),
\end{equation}
where $\Pi(\phi_A^{\rm d})$ and $\Pi(\phi_A^{\rm b})$ denote the osmotic pressure in the droplet and in the bulk, respectively, by recalling $\Pi(\phi_A)=-f_{\rm mix}(\phi_A)+\phi_Af'_{\rm mix}(\phi_A)$.
Eq.~(\ref{mechanic}) indicates the mechanic balance between the osmotic pressure, the surface tension, and the elastic pressure.
The inclusion of the elastic contribution might induce different growth of the droplets than the standard Ostwald ripening \cite{Rosowski2020}.
%Essentially the liquid-liquid phase separation occurs when the volume fraction of droplets $\nu>0$ with $\phi_A^{\rm d} \neq \phi_{A}^{\rm b}$.

\paragraph{Droplet size and density.}
Droplets rich in either \emph{A} or \emph{B}-component will form when liquid-liquid phase separation occurs.
After assuming that the droplets are monodisperse, the size of droplets can be obtained by optimizing the total energy density with respect to the radius of phase separated droplets $R$, $\partial\mathcal{F}/\partial R = - 3\, \nu \, \gamma/R^2 + \nu\, \ud f_{\rm el}/\ud R = 0$, following
\begin{equation}
  \label{R}
  \frac{3\gamma}{R^2} = \frac{\ud f_{\rm el}}{\ud R} \, .
\end{equation}
Eq.~(\ref{R}) indicates that the size of the droplets is determined by the competition between the surface tension of the droplets and the elastic contribution of deformed elastic medium.
By introducing the ratio between the droplet radius and the mesh size of the network, $\lambda_R=R/R_{0}$, we can rewrite Eq.~(\ref{R}) as
$3 {\rm Ec_T} =\lambda_R^2/G \cdot \ud f_{\rm el}/\ud \lambda \big|_{\lambda = \lambda_R}$ with the \emph{thermal elasto-capillary number} defined as $\mathrm{Ec_T}= \gamma/(GR_0)= \gamma /[G^{2/3} (k_BT)^{1/3}]$, where the mesh size $R_0\simeq (k_{\mathrm{B}}T/G)^{1/3}$ is taken.
Alternatively, we can define a function $g(\lambda_R)$~\cite{suppl}:
\begin{equation}
  g(\lambda_R) = \frac{\lambda_R^2}{G} \, \frac{\ud f_{\rm el}}{\ud \lambda} \Big|_{\lambda=\lambda_R}
    = \frac{ 3 \lambda_R^4 }{G (\lambda_R^3 -1 )^2 } W(\lambda_R) .
\end{equation}
Once we know the thermal elasto-capillary number ${\rm Ec_T}$ (given by the surface tension and the shear modulus), then the droplet radius can be obtained as:
$\lambda_R = g^{-1}(3{\rm Ec_T})$,
$R = R_0 \, g^{-1} (3{\rm Ec_T})$.
%Note that the droplet radius is determined only by the surface tension, $\gamma$, and the shear modulus of the elastic network, $G$.

\begin{figure*}[htbp]
\centering
\includegraphics[width=2\columnwidth]{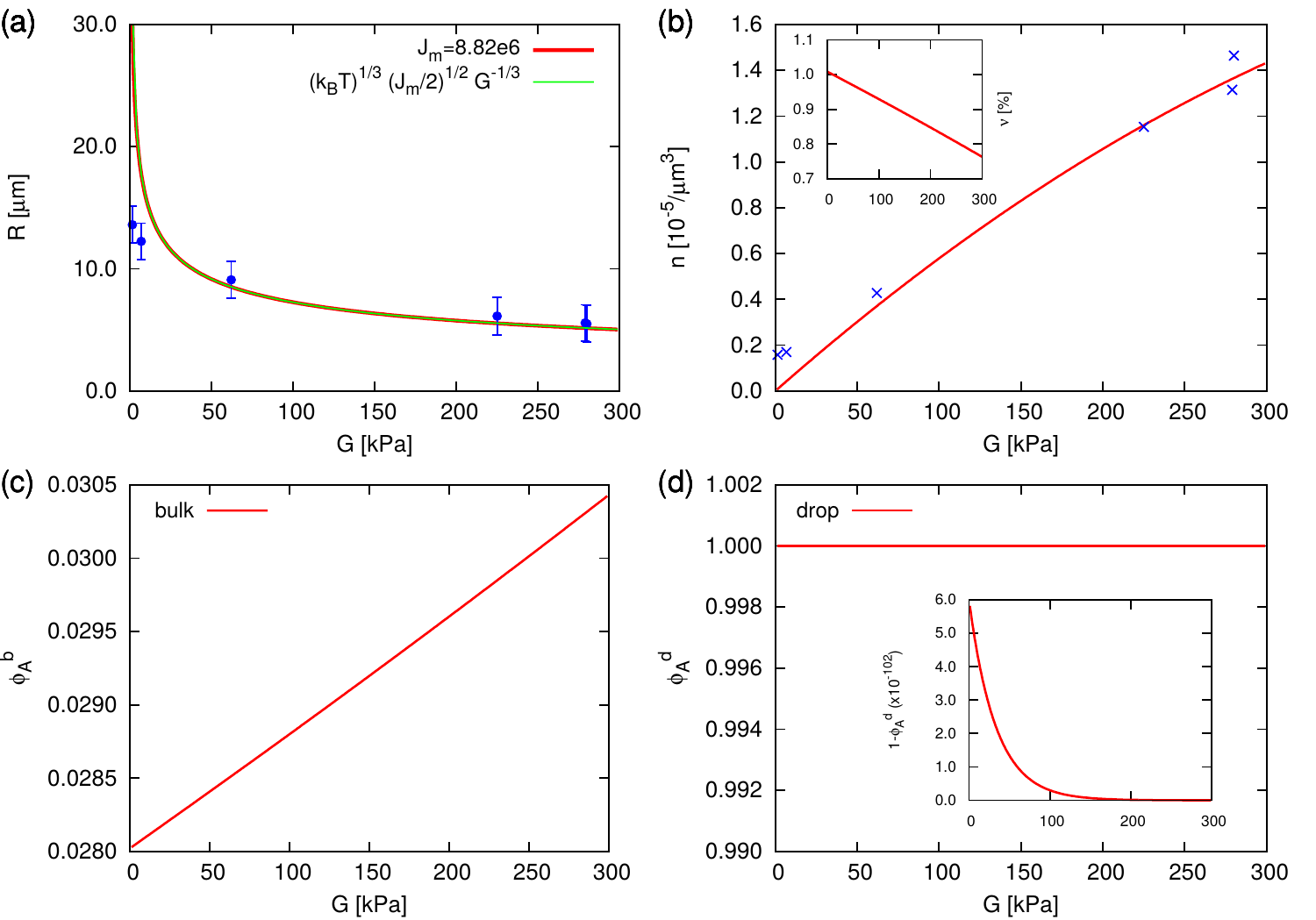}
\caption{(a) Droplet radius, (b) number density, (c) volume fraction of fluorinated oil (\emph{A}-component) in the bulk $\phi_{A}^{b}$ and (d) volume fraction of fluorinated oil in the droplets $\phi_{A}^{d}$ as a function of the shear modulus of the elastic network. The lines correspond to theoretical results and the dots are experimental data~\cite{Style2018}. Parameters are taken as: volume fraction of fluorinated oil at the prepared state (before phase separation) $\phi_A^{0}=0.038$ (saturation volume fraction $\phi_{\mathrm{sat}}$ at $45^{\circ}\mathrm{C}$, volume of one fluorinated oil molecule $v_A= 3.7\times 10^{-28}\mathrm{m}^3$, volume of one silicone oil molecule $v_B= 4.8\times 10^{-26}\mathrm{m}^3$, surface tension $\gamma=0.004 \, \mathrm{N}/\mathrm{m}$, the Flory-Huggins parameter $\chi=2.763$, and $J_{\rm m} = 8.82 \times 10^6 $ in the Gent model.}
\label{f2}
\end{figure*}

With the Gent model, the ratio of the droplet radius to the mesh size of the network, $\lambda_R$, increases with ${\rm Ec_T}$ and soon approaches its limiting value $\lambda_{\rm m}$ where $\lambda_{\rm m}\rightarrow \sqrt{J_{\mathrm{m}}/2}$ for $ J_{\mathrm{m}}\gg1$.
In the range of ${\rm Ec_T}> 5$, the approximated solution is:
\begin{equation}
  \label{eq:R_app}
  \lambda_R \simeq \lambda_{\rm m}, \quad R \simeq \lambda_{\rm m} R_0,
\end{equation}
meaning that the size of the droplet is essentially determined by the limiting stretchability and the mesh size of the material if ${\rm Ec_T}$ is sufficiently large.
One can show that the implementation of the finite stretchability in Gent model is very important in predicting the droplet size, while elastic models without this implementation, \emph{e.g}., neo-Hookean model, does not work \cite{suppl}.
For Gent model with $\lambda_{\rm m}>100$,
the pressure difference between the droplet and the bulk,
$\Delta P$ in Eq.~(\ref{mechanic}),
is almost a constant:
$\Delta P \simeq 5 G/2$~\cite{Gent1969,Gent1991}.

\paragraph{Comparison with experiment.}
In the experiments \cite{Style2018},
a fluorinated oil-silicone oil mixture is first prepared in a silicone polymer network at $45^{\circ}\mathrm{C}$,
with the fluorinated oil at its saturation volume fraction, \emph{i.e.}, $\phi_{\mathrm{sat}}(45^{\circ}\mathrm{C})\sim 0.038$.
(saturation volume fraction $\phi_{\mathrm{sat}}$: critical volume fraction where a free binary fluid mixture without elastic network can phase separate).
Note that due to the presence of the elastic network, there is no phase separation of this fluorinated oil-silicone oil mixture here.
Then by quenching the system to $23^{\circ}\mathrm{C}$,
at which a free mixture would have a lower saturation volume fraction [$\phi_{\mathrm{sat}}(23^{\circ}\mathrm{C})\sim 0.028$], phase separation takes place with the formation of micro-sized droplets rich in fluorinated oil.
By taking the \emph{A}- and \emph{B}- component in the model as the fluorinated oil and the silicone oil, respectively,
we can directly compare the theoretical results with the experiments.
In the experiments~\cite{Style2018}, $v_A\sim 3.7\times 10^{-28}\mathrm{m}^3$ and $v_B\sim 4.8\times 10^{-26}\mathrm{m}^3$.
The surface tension is $\gamma\sim0.004 \, \mathrm{N}/\mathrm{m}$ and the Flory-Huggins parameter at 23$^{\circ}$C is $\chi=2.763$.
In this particular experiment,
although surface tension is important in determining the size of phase separated droplets as shown in Eq.~(\ref{R}),
its contribution is negligible in the criterion of phase separation compared with other terms (osmotic pressure and elastic contribution) in Eq.~(\ref{mechanic}).

\begin{figure}[htbp]
\includegraphics[width=1.1\columnwidth]{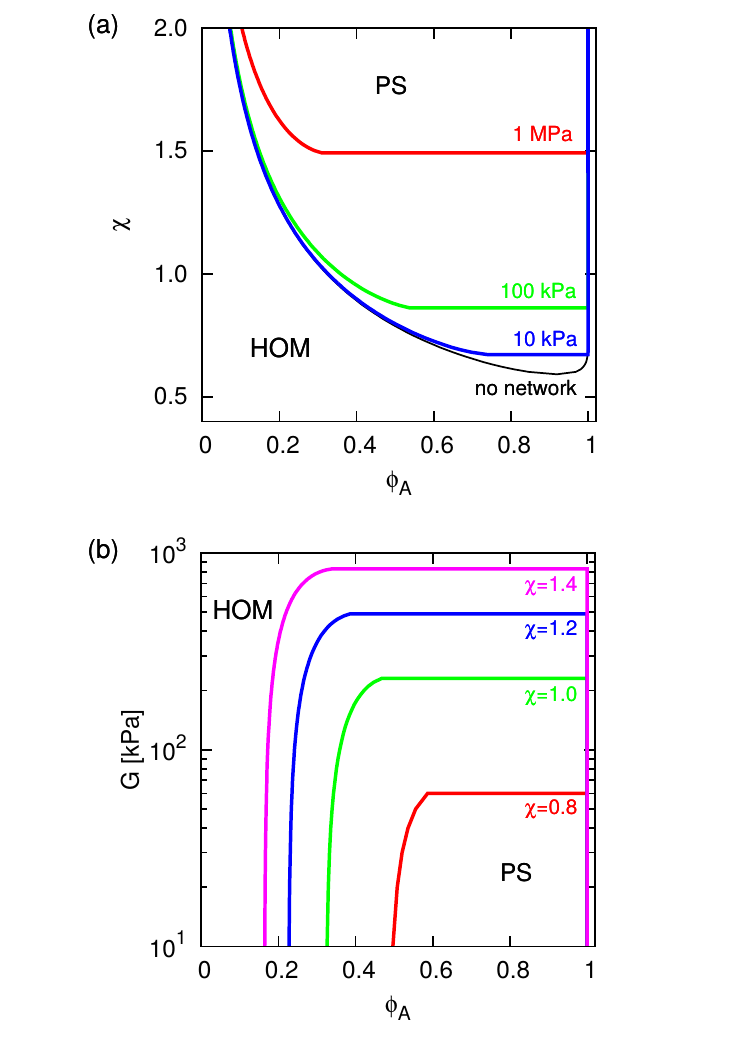}
\caption{Phase diagram of elastically mediated liquid-liquid phase separation (a) in the plane of ($\phi_{A},\chi$) for different shear modulus $G$ and (b) in the plane of ($\phi_{A}, G$) for different Flory-Huggins parameter $\chi$; HOM denotes homogeneous state without phase separation and PS denotes phase separation. Parameters are taken as: volume of one fluorinated oil molecule $v_A = 3.7\times 10^{-28}\mathrm{m}^3$, and volume of one silicone oil molecule $v_B = 4.8\times 10^{-26}\mathrm{m}^3$.}
\label{fig3}
\end{figure}

As shown in Fig.~\ref{f2}(a), the droplet size decreases with
the shear modulus of the elastic network, matching with the experimental observations, and there is a relation between the droplet size and shear modulus: $R\simeq(k_{\mathrm{B}}T)^{1/3} (J_{\mathrm{m}}/2)^{1/2} G^{-1/3}$,
following the analytic expression in Eq.~(\ref{eq:R_app}).
Note that the stretching limit $J_{\mathrm{m}}= 8.82 \times 10^{6}$ is taken as a constant independent of the shear modulus.
One can also introduce the dependence of $J_{\mathrm{m}}$ on the shear modulus and can easily obtain a better fitting in Fig.~\ref{f2}(a),
but here we will keep $J_{\mathrm{m}}$ as a constant for simplicity.
As noted in the inset of Fig.~\ref{f2}(b),
the volume fraction of droplets $\nu$ slightly decreases by changing the shear modulus, in the form of $\nu=- 8 \cdot 10^{-6}~ \mathrm{kPa}^{-1} \cdot G +0.01$ (changing from $\nu\simeq1\%$ without elastic network to $\nu\simeq0.76\%$ in a stiff network with $G=300\, \mathrm{kPa}$).
Meanwhile, the number density of the droplets can be obtained as $n=3\nu/(4 \pi R^{3})\simeq 6 \sqrt{2}\nu \, G/\big(4 \pi k_{\mathrm{B}}TJ_{\mathrm{m}}^{3/2}\big)$, which basically increases linearly with the shear modulus, $n\sim G$ for small $G$ as shown in Fig.~\ref{f2}(b).
In other words, more and smaller droplets will form if the elastic network is stiffer.
Remarkably, when phase separation occurs, the volume fraction of fluorinated oil (\emph{A}-component) in the droplets $\phi_{A}^{d}$ remains almost as a constant near \emph{1} regardless of the shear modulus $G$ in Figure~\ref{f2}(d), while its volume fraction in the bulk increases almost linearly with the shear modulus, $\phi_{A}^{b}(G)\simeq\phi_{A}^{b}(0)-\nu=0.028+ 8 \cdot 10^{-6}~ \mathrm{kPa}^{-1} \cdot G$ as shown in Fig.~\ref{f2}(c);
this means that the droplets consist of almost pure fluorinated oil.

In experiments, the shear modulus of the elastic network is taken as a controllable parameter, where such elastic constraint can control the size of phase separated droplets as discussed above.
Meanwhile, the elastic constraint can also inhibit the liquid-liquid phase separation.
Figure \ref{fig3}(a) shows the phase diagram in the $\phi_A$--$\chi$ plane for different shear modulus $G$.
At small $\chi$, the system is homogeneous.
When the network is absent, the system is a binary mixture of $A$/$B$ with a aspect ratio $N=v_B/v_A$.
There is a critical point at $(\phi_A)_c = 1/(\sqrt{N}+1)$, $\chi_c = {1}/{2} + {1}/{\sqrt{N}} + 1/(2N)$ \cite{RubinsteinColby}.
When $\chi > \chi_c$, there is a certain regime of $\phi_A$ at which the homogeneous state becomes unstable and phase separation proceeds.
The presence of the elastic network moves the binodal line upward in the small-$\phi_A$ region and changes the critical point to a critical line.
Therefore, there are regions between the colored lines and the black line in Fig.~\ref{fig3}(a) that become stable and remain homogeneous due to the elastic network.
The critical value of the Flory-Huggins parameter $\chi_c$ increases with the increasing shear modulus $G$.
This is more clear in Fig.~\ref{fig3}(b), which shows the phase diagram in the $\phi_A$--$G$ plane for different values of $\chi$.
Initially at small value of $G$, there is a range of $\phi_A$ that homogeneous states are unstable.
Increasing the shear modulus will slight reduce the unstable region, and when certain critical value of $G_c$ is reached, the phase-separated region disappears and the homogeneous state becomes stable.

In conclusion, we propose a continuum model of liquid-liquid phase separation in an elastic network incorporating fluids mixing, droplet nucleation, network deformation, thermodynamic fluctuation, \emph{etc}., and investigate quantitatively how network elasticity can influence such phenomenon.
When phase separation occurs, the size of the phase separated droplets is found to decrease with the increasing network elasticity in the form of a scaling law, and the number density of the droplets increases almost linearly;
the theoretical results are also verified by the experimental observations.
Furthermore, phase diagrams in the planes of Flory-Huggins parameter, shear modulus, and the volume fraction of liquid compositions are constructed, providing necessary conditions for liquid-liquid phase separation in an elastic medium.
We believe that this portable model helps to understand the elastically mediated liquid-liquid phase separation, and can also be predictive of how to produce phase separated droplets of desired size and density.

\begin{acknowledgments}
This work is supported by the Strategic Priority Research Program of Chinese Academy of Sciences (Grant No. XDA17010504) and the National Natural Science Foundation of China (Grant No. 11947302 and 21774004).
F. M. acknowledges partial funding from Alexander von Humboldt Foundation. The computations of this work were conducted on the HPC cluster of ITP-CAS.
\end{acknowledgments}

\end{document}

\title{Supplementary Information:\\
{\it Modelling Elastically-Mediated Liquid-Liquid Phase Separation}}

\author{Xuefeng Wei}
\thanks{X. W. and J. Z. contributed equally to this work. }
\affiliation{CAS Key Laboratory of Theoretical Physics, Institute of Theoretical
Physics,Chinese Academy of Sciences, Beijing 100190, China}
\affiliation{School of Physical Sciences, University of Chinese Academy of Sciences, China}

\author{Jiajia Zhou}
\thanks{X. W. and J. Z. contributed equally to this work. }
\affiliation{Key Laboratory of Bio-Inspired Smart Interfacial Science and Technology of Ministry of Education, School of Chemistry, Beihang University, Beijing 100191, China}
\affiliation{Center of Soft Matter Physics and Its Applications, Beihang University, Beijing 100191, China}

\author{Yanting Wang}
\affiliation{CAS Key Laboratory of Theoretical Physics, Institute of Theoretical
Physics,Chinese Academy of Sciences, Beijing 100190, China}

\author{Fanlong Meng}
\email{fanlong.meng@itp.ac.cn}
\affiliation{CAS Key Laboratory of Theoretical Physics, Institute of Theoretical
Physics,Chinese Academy of Sciences, Beijing 100190, China}

\maketitle

\beginsupplement

\section{More details of function $\boldsymbol{g(\lambda_{R})}$}
The elastic energy density by the droplets deforming the elastic medium:
\begin{equation}
  f_{\rm el}(R) = 3 \left[ 1 -  \left( \frac{R_0}{R} \right)^3 \right]
                    \int_1^{R/R_0} \frac{\lambda^2 W(\lambda) }{(\lambda^3-1)^2} \ud \lambda ,
  \label{eq:fel1}
\end{equation}
where $W(\lambda)$ in the elastic term represents the elastic energy density of an inflated spherical shell (like a spherical balloon).
The size of droplets follows the relation:
\begin{equation}
  \label{eq:R1}
  \frac{\lambda_R^2}{G} \frac{\ud f_{\rm el}}{\ud \lambda} \Big|_{\lambda = \lambda_R}
  = \frac{3\gamma}{G R_0}
  = 3 {\rm Ec_T}.
\end{equation}
The right-hand side of Eq.~(\ref{eq:R1}) is a constant for given system, where ${\rm Ec_T}=\gamma/(G R_0)$ denotes the \emph{thermal elasto-capillary number}.
The left-hand side of Eq.~(\ref{eq:R1}) is a function of $\lambda_R$,
for which we define $g(\lambda_{R})$,
\begin{equation}
  \label{eq:R2}
  g(\lambda_R)=\frac{\lambda_R^2}{G} \frac{ \ud {f}_{\rm el}}{\ud \lambda} \Big|_{\lambda = \lambda_R}= \frac{ 3 \lambda_R^4 }{G (\lambda_R^3 -1 )^2 } W(\lambda_R).
\end{equation}
Then Eq.~(\ref{eq:R1}) can be re-written as:
\begin{equation}
  g(\lambda_R) = 3 {\rm Ec_T}.
\end{equation}

%-------------------------------
\begin{figure}[htbp]
  \includegraphics[width=0.6\columnwidth]{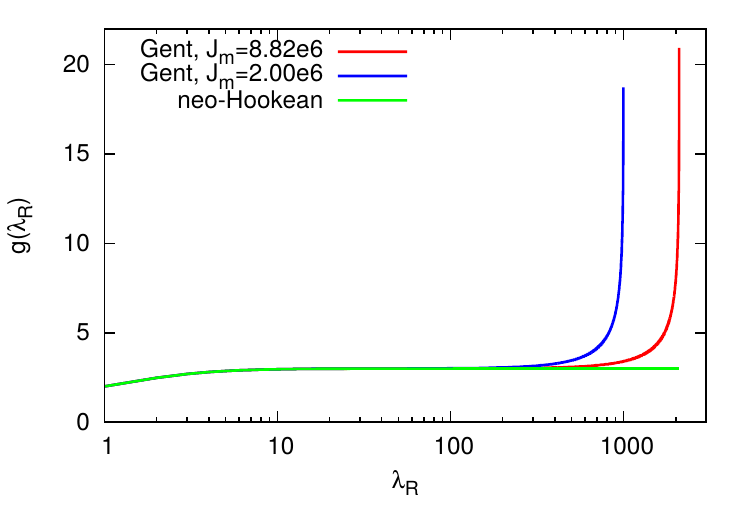}
  \caption{Function $g(\lambda_R)$ for neo-Hookean model and Gent model.}
  \label{fig:S1}
\end{figure}
%-------------------------------

\emph{Neo-Hookean model}. For neo-Hookean model, $W(\lambda_R)=G(2\lambda_R^{2}+\lambda_R^{-4}-3)/2$, $g(\lambda_R)$ is:
\begin{equation}
 g(\lambda_R) = \frac{ 3 \lambda_R^4}{ (\lambda_R^3 -1)^2 }
    \frac{1}{2} (2\lambda_R^2 + \lambda_R^{-4} - 3)
\end{equation}
of which the value changes from $2$ at $\lambda_R\rightarrow1$ to $3$ at $\lambda_R\rightarrow\infty$.
In other words, the droplet size can only be calculated by neo-Hookean model for a very narrow range of the thermal elasto-capillary number,  $2/3<{\rm Ec_T}<1$.

\emph{Gent model}.
For Gent model, $W(\lambda_R)=-\frac{1}{2} G J_{\rm m} \ln [1 - {(2\lambda_R^{2} + \lambda_R^{-4} - 3)}/{J_{\rm m}}]$ with $J_{\rm m}$ characterizing the finite stretchability of the material, $g(\lambda_R)$ is:
\begin{equation}
 g(\lambda_R) = -\frac{ 3 J_{\mathrm{m}}\lambda_R^4 }{2 (\lambda_R^3 -1 )^2}
 \ln \left[ 1-\frac{2\lambda_R^{2}+\lambda_R^{-4}-3}{J_{\rm m}} \right]
\end{equation}
which reduces to that of neo-Hookean model if the deformation is small and diverges if $\lambda_R \rightarrow \lambda_{\rm m} \simeq \sqrt{J_{m}/2}$ as shown in Fig.~\ref{fig:S1}.

\section{Droplet size as a function of $\boldsymbol{{\rm Ec_T}}$}
\begin{figure}[htbp]
  \includegraphics[width=0.6\columnwidth]{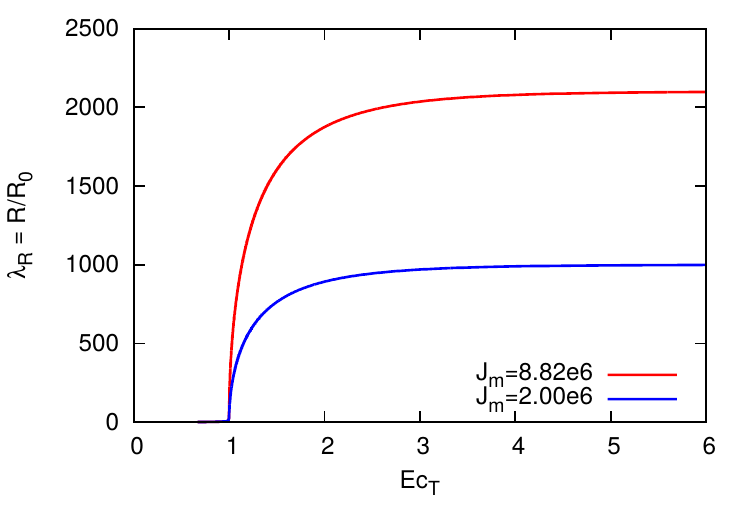}
  \caption{Function $g^{-1}({\rm Ec_T})$ for Gent model of $J_{\rm m}=2.00 \times 10^{6}$ and $8.82 \times 10^{6}$.}
  \label{fig:S2}
\end{figure}

The ratio of the droplet radius over the mesh size, $\lambda_{R}=R/R_{0}$, depends on the thermal elasto-capillary number, ${\rm Ec_T}$, in the form of:
\begin{equation}
  \lambda_R = g^{-1}(3{\rm Ec_T}), \quad
  R = R_0 g^{-1} (3{\rm Ec_T}),
\end{equation}
which is plotted in Figure~\ref{fig:S2}.

%--------------------------------------
%--------------------------------------

\section{Pressure difference as a function of $\boldsymbol{\lambda_{R}}$. }

\begin{figure}[htbp]
  \includegraphics[width=0.6\columnwidth]{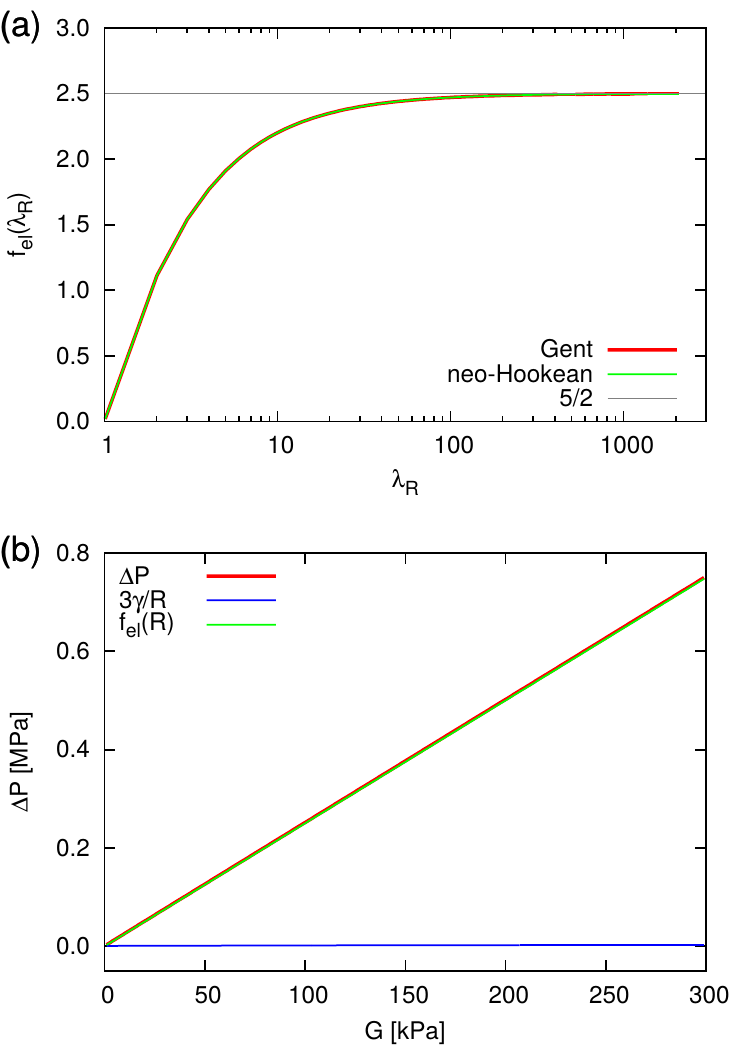}
  \caption{(a) Function ${f}_{\rm el}(\lambda_R)$. The Gent model ($J_{\rm m}=8.82 \times 10^{6}$ or $\lambda_{\rm max}=2100$) only differs from the neo-Hookean model at large stretching ratios. (b) The pressure difference $\Delta P$ as a function of $G$. The contributions from the interfacial tension and the elastic network are plotted in blue and green, respectively.}
  \label{fig:S3}
\end{figure}

The pressure difference defined in the main text as $\Delta P = \Pi(\phi_A^d)-\Pi(\phi_A^b)$ can be calculated as:
\begin{equation}
  \label{eq:DeltaP_app}
  \Delta P = \frac{3\gamma}{R}+ f_{\rm el}(R) = G\left[ \frac{3 {\rm Ec_T}}{\lambda_R} + \frac{f_{\rm el}(\lambda_R)}{G} \right].
\end{equation}
The elastic energy $f_{\rm el}(\lambda_R)$ for Gent model and neo-Hookean model are plotted in Fig.~\ref{fig:S3}(a).
For typical experimental values \cite{Style2018}, the elastic contribution in (\ref{eq:DeltaP_app}) dominates over the interfacial tension contribution.
This is shown in Fig.~\ref{fig:S3}(b).

%--------------------------------
%-------------------------------

\section{Figure 2(a) with $\boldsymbol{G}$-dependent $\boldsymbol{J_{\rm m}}$}

\begin{figure}[htbp]
  \includegraphics[width=0.6\columnwidth]{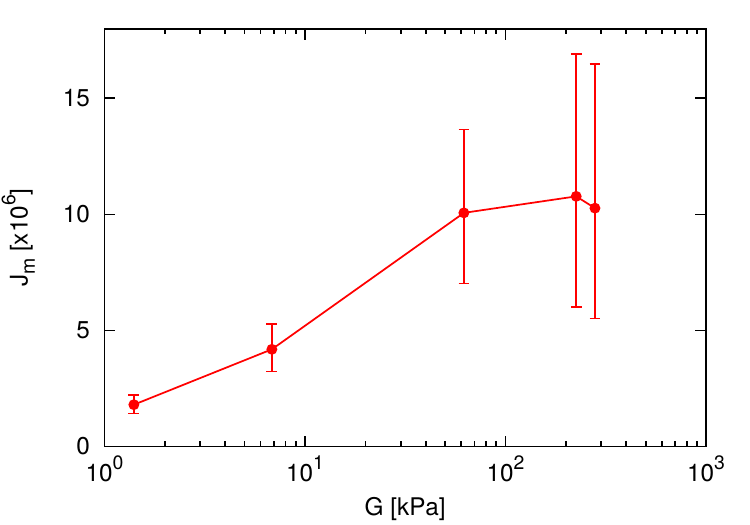}
  \caption{The dependence of $J_{\rm m}$ on the shear modulus $G$. }
  \label{fig:S4}
\end{figure}

Rather than taking a constant $J_{\rm m}$ for different shear modulus in Figure 2(a) of the main text, one can adapt a $G$-dependent $J_{\rm m}(G)$ for a perfect fitting.
The dependence of $J_{\rm m}$ on the shear modulus $G$ is shown in Fig.~\ref{fig:S4}, where $J_{\rm m}$ increases with the shear modulus $G$.

%--------------------------------
%--------------------------------

\bibliography{ref.bib}